# The Search for Extraterrestrial Intelligence in Earth's Solar Transit Zone

René Heller[1,2] and Ralph E. Pudritz[2,3,4,5]

## Abstract

Over the past few years, astronomers have detected thousands of planets and candidate planets by observing their periodic transits in front of their host stars. A related method, called transit spectroscopy, might soon allow studies of the chemical imprints of life in extrasolar planetary atmospheres. Here, we address the reciprocal question, namely, from where is Earth detectable by extrasolar observers using similar methods. We explore Earth's transit zone (ETZ), the projection of a band around Earth's ecliptic onto the celestial plane, where observers can detect Earth transits across the Sun. The ETZ is between 0.520° and 0.537° wide due to the non-circular Earth orbit. The restricted ETZ (rETZ), where Earth transits the Sun less than 0.5 solar radii from its center, is about 0.262° wide. We first compile a target list of 45 K and 37 G dwarf stars inside the rETZ and within 1 kiloparsec (about 3260 lightyears) using the Hipparcos catalog. We then greatly enlarge the number of potential targets by constructing an analytic galactic disk model and find that about $10^5$ K and G dwarf stars should reside within the rETZ. The ongoing GAIA space mission can potentially discover all G dwarfs among them (several $10^4$) within the next five years. Many more potentially habitable planets orbit dim, unknown M stars in the ETZ and other stars that traversed the ETZ thousands of years ago. If any of these planets host intelligent observers, they could have identified Earth as a habitable, or even as a living, world long ago, and we could be receiving their broadcasts today. The K2 mission, the Allen Telescope Array, the upcoming Square Kilometer Array, or the Green Bank Telescope might detect such deliberate extraterrestrial messages. Ultimately, the ETZ would be an ideal region to be monitored by the Breakthrough Listen Initiatives, an upcoming survey that will constitute the most comprehensive search for extraterrestrial intelligence so far.

## 1. Introduction

Now that thousands of extrasolar planets have been discovered in a small part of the sky by the Kepler space telescope (Batalha et al., 2013; Lissauer et al., 2014; Rowe et al., 2014), and Earth-sized habitable-zone (HZ) planets turn out to orbit about one out of ten stars (Kaltenegger and Sasselov, 2011; Dressing and Charbonneau, 2013; Petigura et al., 2013), the search for life outside the Solar System has experienced a substantial impetus. Recent studies suggest that deciphering the bio-relevant constituents from Earth-like planetary atmospheres – such as oxygen, water, carbon dioxide, and methane – is feasible by taking high-accuracy transit transmission spectra of these world's atmospheres (Ehrenreich et al., 2006; Kaltenegger and Traub, 2009; Rauer et al., 2011; Belu et al., 2011; von Paris et al., 2011; Benneke and Seager, 2012). In addition to other techniques such as spatially resolved spectroscopy from space (Leger et al., 2015), transit observations are thus one key technology with which to characterize extrasolar, inhabited worlds.

In the spirit of Carl Sagan's view of Earth as "a pale blue dot" from the perspective of an observer on the receding Voyager 1 spacecraft (Sagan, 1994), we propose a similar, but more probing, question. Supposing that extraterrestrial observers are using similar transit techniques and are also surveying the galaxy for habitable worlds with atmospheres that are obviously altered by life: Which of them could discover Earth as a planet that supports life?

The planetary systems, from where it is possible to detect Earth transiting in front of the Sun, are all located in a thin strip around the ecliptic as projected onto the galaxy, which we refer to as Earth's transit zone (ETZ). Possible extraterrestrials that observe Earth from within the ETZ will have direct evidence to classify Earth as a living planet perhaps worth the effort of deliberate broadcasts. Hence, stellar systems within the ETZ are promising targets for us to search for extraterrestrial intelligence (SETI).

More than 100 SETI surveys have been performed in the past century (Tarter, 2001). Searches have focused

[1] Max Planck Institute for Solar System Research, Justus-von-Liebig-Weg 3, 37077 Göttingen, Germany; heller@mps.mpg.de
[2] Origins Institute, McMaster University, 1280 Main Street West, Hamilton, ON L8S 4M1, Canada
[3] Department of Physics and Astronomy, McMaster University; pudritz@physics.mcmaster.ca
[4] Max Planck Institute for Astronomy, Königstuhl 17, 69117 Heidelberg, Germany
[5] Heidelberg University, Center for Astronomy, Institute for Theoretical Astrophysics, Albert-Ueberle-Str. 2, 69120 Heidelberg, Germany

either on nearby systems (Turnbull and Tarter, 2003a), mostly because an extrasolar signal emitted with a given power is stronger and more likely to be detected at closer distances, or on stellar clusters (Turnbull and Tarter, 2003b) where potential messages from many objects can be collected in a single observational run. The first targeted SETI was project OZMA in 1960, which used the 26 m radio telescope of the National Radio Astronomy Observatory. One of the most popular surveys is SETI@home (Korpela et al., 2001), which digests data of the 300 m Arecibo radio telescope. Project Phoenix (Backus & Project Phoenix Team 2001), which was conducted by the SETI Institute at different sites between 1995 and 2004, is one of the most ambitious searches ever undertaken, covering more than 800 stars as far as 240 lightyears (lys), or about 74 parsec (pc), from Earth. More recent searches include the first very long baseline interferometric SETI (Rampadarath et al., 2012), placing an upper limit of 7 MW $\mathrm{Hz}^{-1}$ on the power output of any isotropic emitter in the planetary system around the M dwarf Gl 581. The Allen Telescope Array (ATA), jointly operated by the SETI Institute and the Radio Astronomy Laboratory (University of California), has been used for targeted surveys around the galactic center (Tarter et al., 2002; Williams et al., 2009). Later, Tellis and Marcy (2015) searched for optical laser emission from more than one thousand Kepler Objects of Interest with the Keck telescope. And most recently, Yuri Milner and Stephen Hawking announced the "Breakthrough Listen Initiatives"[1], a new effort involving a 100 Million USD investment over the next ten years to search for extraterrestrial broadcasts using both radio and optical wavelengths (Merali, 2015).

Major obstacles for SETI include the questions of where to point, in which frequency range to search, and which sensitivity be required. If messages from other observers were transmitted to us intentionally (so-called "beacons"), then their power could be relatively strong. However, if their broadcasts reach us unintentionally (by "leakage"), they might be very weak and short-lived (Forgan and Nichol, 2011). On the other hand, these leakage messages could be much more abundant in the Universe, because beacons would require the emitter (1) to have an interest in interstellar communication as well as (2) the capabilities to do so, and (3) they would need to actually identify Earth as a worthwhile target. Ultimately, while leakage transmissions should preferably reach us from regions with higher stellar (more precisely: planetary) densities, simply because we should expect more emitters where there are more potentially habitable worlds, a preferred celestial region for intended messages has not been agreed upon. Here, we suggest that this preferred region is the ETZ and present a list of targets to listen for intended transmissions.

The idea to search for intended messages within the ETZ is not new (Filippova and Strelnitskij, 1988), but it has never been treated in the refereed literature. Filippova (1990) was one of the first to present a list of nearby stars (29 most KGF dwarfs) near the ecliptic to be monitored, and Castellano et al. (2004) presented a list of 17 G dwarfs from the Hipparcos catalog (ESA, 1997). The latter report was reconsidered by Shostak & Villard (2004), who proposed to survey the Sun's anti-direction for broadcasts and utilize Earth's solar transit as a means of synchronization between potential senders and us. Conn Henry et al. (2008) proposed a near-ecliptic SETI survey using the ATA. A broader approach was taken by Nussinov (2009), who proposed that solar transits of any of the solar system planets might draw the interest of extraterrestrials. With the repurposed K2 mission now monitoring thousands of stars near the ecliptic, with the SKA starting observations within a few years from now, and with the Breakthrough Listen Initiative about to be launched, a detailed and updated treatment of the ETZ is now timely.

Here, we present a more rigorous geometric description of the ETZ (Sect. 2) than given in the abovementioned studies. We almost triple the list of target stars for SETI in the ETZ (Sect. 3.1), and we compare the number of stars on that target list with the number of stars actually expected in the ETZ and within 1 kiloparsec (kpc)[2] to the Sun, based on stellar density measurements (Sect. 3.2).

## 2. Earth's Transit Zone

### *2.1 Geometrical construction*

We start our geometrical construction of the ETZ by assuming that Earth's solar orbit is circular. For extrasolar observers to see Earth completely in transit, as opposed to a grazing transit, they must see the whole planetary disk passing in front of the Sun. Such a full transit can be observed from inside a zone with an angular size $\varphi^{\mathrm{ETZ}} = 2(\alpha - \gamma)$, where $\alpha = \arctan(R_{\odot}/a)$, $a = 1$ AU, and $\gamma = \arcsin(R_{\oplus}/t)$ (Fig. 1). Using Pythagoras' theorem for $t$ as the hypotenuse of the bold triangle in Fig. 1, we obtain $t = (a^2 + R_{\odot}^2)^{1/2}$ and

$$\varphi^{\rm ETZ} = 2\left(\arctan\left(\frac{R_\odot}{a}\right) - \arcsin\left(\frac{R_\oplus}{\sqrt{a^2+R_\odot^2}}\right)\right) \quad . \tag{1}$$

[1] www.breakthroughinitiatives.org

[2] 1 kiloparsec corresponds to about 3,262 lightyears or $3.1 \times 10^{19}$ m.

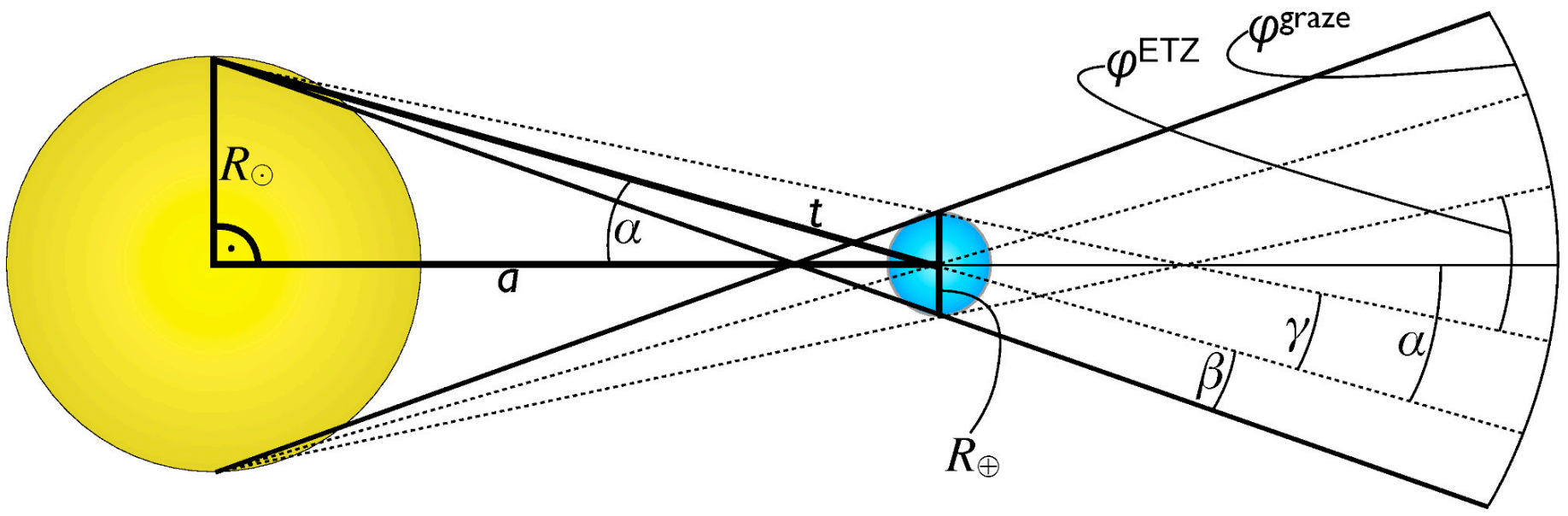

**FIG. 1** Geometrical construction of Earth's transit zone (ETZ). The yellow circle represents the Sun, the blue circle Earth (sizes not to scale). Within $\varphi^{\rm ETZ}$, the whole disk of Earth performs a transit while at $\varphi^{\rm graze}$ the Earth scrapes through a grazing transit. The projection of the stripe defined by $\varphi^{\rm ETZ}$ on the sky defines the ETZ around the Sun.

With $R_\oplus \ll R_\odot$ and $\arctan(x) \approx x$ for $x \rightarrow 1$, Eq. (1) becomes $\varphi^{\rm ETZ} \approx R_\odot/a$, as used by Castellano et al. (2004). While these authors compute $\varphi^{\rm ETZ} \approx 2 \times 0.267° = 0.534°$, our Eq. (1) yields $\varphi^{\rm ETZ} \approx 0.528°$. This result is also somewhat more precise than the approximation of $2 \times 0.28° = 0.56°$ by Nussinov (2009). As an aside, the somewhat wider celestial band spanned by $\varphi_{\rm graze} = 2(\alpha+\beta)$, with $\beta = \arctan(R_\oplus/t)$, defines the outermost zone from which a grazing transit can be seen, and it has a width of 0.538°.

To estimate how the area within the ETZ ($A_{\rm ETZ}$) compares to the total area of the celestial plane ($A_{\rm tot}$), note that $A_{\rm tot} = 4\pi$, while

$$A_{\rm ETZ} = \int_0^{2\pi} {\rm d}\phi \int_{\pi/2-\varphi^{\rm ETZ}/2}^{\pi/2+\varphi^{\rm ETZ}/2} {\rm d}\theta \ \sin\theta \approx \ 4.608 \times 10^{-3} A_{\rm tot} \quad . \tag{2}$$

Constraining SETI to the ETZ thus implies a substantial reduction of the celestial area to be surveyed.

The transit impact parameter $b$, which describes the minimum separation of the crossing planet from the stellar center in units of stellar radii, is crucial for an observer. For grazing transits ($b \rightarrow 1$), the transit duration $D$ can be very short, making characterization of the planetary atmosphere extremely challenging, because even an extraterrestrial observer would be confronted with the solar photometric variability. It is thus plausible to constrain the search window to stars, from which they would watch Earth's solar transit with $b \leq 0.5$. Replacing $R_\odot$ in Eq. (1) with $R_\odot/2$ yields a restricted ETZ (rETZ) that is 0.262° wide.

With Earth's orbital mean motion ($\omega$) of about 360°/365.25d ≈ 0.0411°/h, the duration $D$ of the Earth transit at $b = 0$ (over the solar diameter) is

$$D \approx \frac{\varphi^{\rm ETZ}}{\omega} \approx 12\,{\rm h} : 51{\rm min} \quad . \tag{3}$$

If the transit were observed with $b \leq 0.5$, then its duration would always be longer than $3^{1/2}\,D/2 \approx 11$ h : 08min – a reasonable duration that might allow extraterrestrials to detect Earth's atmospheric biomarkers.

### *2.2 Variation of the Earth's transit zone*

Various physical effects complicate the picture shown in Fig. 1, such as

– Earth's variable orbital eccentricity ($e$)
– Earth's apsidal precession
– perturbations of Earth's orbit from the Moon
– solar motions due to planetary perturbations (mostly from Jupiter)
– the precession of the ecliptic

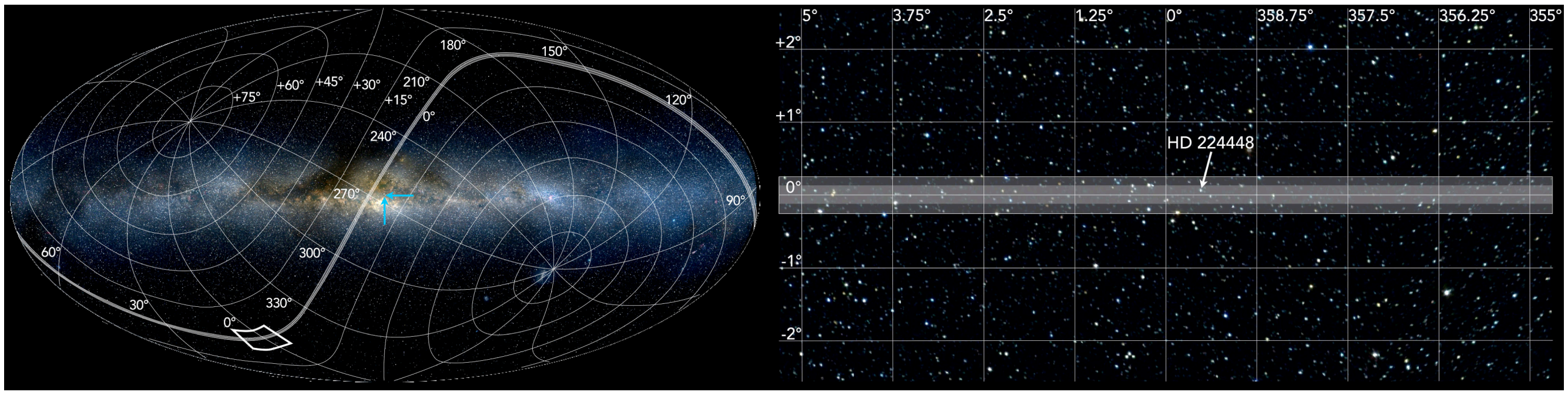

**FIG. 2** Projection of the Earth's transit zone (ETZ) onto the celestial plane. Left panel: All-sky view in a galactic reference frame with ecliptic coordinates, based on optical, infrared (2MASS), and radio (Planck) data. The ETZ is indicated as a gray band curved along the ecliptic. The quadrangle around (0°,0°) indicates the zoom shown in right panel. Two blue arrows label the origin of the galactic coordinate frame. Right panel: Zoom into the region around (0°,0°), now in an ecliptic reference frame. The wide, light shaded region has a width of 0.528°, following Eq. (1), while the dark shaded region highlights the restricted ETZ with a width of 0.262°, where Earth transits the Sun with an impact parameter $b \leq 0.5$. The position of one ETZ G dwarf is indicated with a black arrow.

all of which we discuss in the following.

Different from what has been assumed in Sect. 2.1, Earth's solar orbit is not circular but has an eccentricity ($e$) of about 0.016. Hence, the distance between Earth and the Sun ($a$) is a function of the geocentric ecliptic longitude $\lambda$. Earth's closest solar approach, or perihelion, occurs at the longitude of the periapsis, currently almost exactly $\lambda = 283°$, (around January 4 in this decade). At that instant, the distance between the Sun and Earth is $r_{\rm min} = a(1 - e) \approx 0.9833$ AU. With the Sun at $\lambda = 283°$, potential observers of Earth's transit would be located in the anti-direction at $\lambda = 103°$. Hence, at $\lambda = 103°$ in the ETZ, $\varphi^{\rm ETZ}{}_{\rm min} = 0.537°$. In turn, at aphelion (currently at $\lambda = 103°$ around July 4), observers at $\lambda = 283°$ see Earth's full solar transit within an angle $\varphi^{\rm ETZ}{}_{\rm max} = 0.520°$.

In the long run, gravitational perturbations on Earth's orbit from the other planets (most of all Jupiter and Venus) cause Earth's eccentricity to vary between almost 0 and about 0.06 (Laskar et al., 1993), with the most relevant beat periods occurring on time scales of hundreds of thousands of years (Laskar et al., 2011a). Hence, some parts of the ETZ were as narrow as 0.498° or as wide as 0.562° during these phases. Orbital perturbations from Ceres and Vesta induce chaos on a similar time scale (Laskar et al., 2011b), preventing an exact reconstruction of an ancient ETZ. As Earth's argument of perihelion also evolves, causing the orbital ellipse to revolve once in 112 kyr, those variations in the width of the ETZ should have been smeared out over the ecliptic longitude.

As for the location of the ETZ, variations of the Sun's position relative to the Solar System barycenter, mostly due to Jupiter's perturbations, have the most relevant effect on b of up to 2.5 %. Earth's movement around the Earth-Moon barycenter and the 5.14° misalignment of the lunar orbital plane to the ecliptic induce additional variations of the Earth's solar transit impact parameter. But the maximum vertical displacement of the Earth due to the Moon is just 423 km, or 0.06 % of the solar radius. Hence, this effect is irrelevant for our considerations. Planetary precession, that is, variations of the ecliptic with respect to the invariable plane due to planetary perturbations are also negligible, since they are about 100 times weaker than lunisolar effects (Williams, 1994).

## 3. Host Stars of Potential Observers in Earth's Transit Zone

### *3.1 SETI targets*

Figure 2 shows the path of the ETZ on the celestial plane. The all-sky view in the left panel is a blend of optical data (Mellinger, 2009), a JHK infrared composite of the Two Micron All Sky Survey (2MASS, Skrutskie et al., 2006), and radio emission between 30 and 70 GHz detected by the Planck satellite (The Planck Collaboration, 2006). The data were retrieved with the Aladin Sky Atlas[3] v7.5 (Bonnarel et al., 2000). The arrows in the center of the left panel indicate the origin of the galactic reference frame. Ecliptic coordinates are overplotted to illustrate the orientation of the ETZ with respect to the galactic plane.

At the Sun's location within the Milky Way, the galactic disk has a width of about 600 pc (Binney and Tremaine, 2008). As the Solar System is tilted against the galactic plane by about 63°, the ETZ's pathlength through the disk is roughly 1 kpc long. This length is similar to the width of the galactic habitable zone (Lineweaver et al., 2004), which extends from about 7.5 to 9 kpc from the galactic center, the Sun being at about

[3] http://aladin.u-strasbg.fr

8.5 kpc (for counter-arguments see Prantzos, 2008). Stars beyond 1 kpc from the Sun may thus not host inhabited planets in the first place. Consequently, we use the SIMBAD Astronomical Database[4] to retrieve all stars within 1 kpc to the Sun that reside in the rETZ, where Earth's $b \leq 0.5$. Our query code reads

```
region(ZONE ecl,0.0 +0.0,360.0d 0.262d) & plx >= 1
```

where the "plx" (for parallax) condition is measured in milliarcseconds and the string "ecl" sets the ecliptic coordinate system. This search yields 234 stars, including 15 M, 67 K, 46 G, 58 F, 29 A, and 13 B stars. There is no O star, and five of the remaining six objects are not classified. Finally, the sample contains a very nearby white dwarf ("van Maanen's star", van Maanen, 1917) at a distance of only 4 pc. Obviously, these counts dramatically underestimate the number of M dwarfs, which are the dominant type of star in the solar neighborhood.

We filter 113 K and G stars from this sample. A more sophisticated approach would make use of the stellar ages (if known) for the remaining K and G stars, as done by Turnbull and Tarter (2003a), since some of these targets might still be very young with little time for the emergence of an intelligent species. Nevertheless, most of these stars should be of similar age as the Sun since they are in the solar neighborhood of the Milky Way. The rejection of giants and subgiants finally leaves us with 45 K and 37 G dwarfs. Table 1 lists their coordinates, proper motions, parallaxes, distances, V magnitudes, and spectral types. Parallaxes are obtained from the Hipparcos catalog (van Leeuwen, 2007), and distances in units of parsecs are calculated as $(\text{parallax})^{-1}$.

Castellano et al. (2004) presented a list of 17 G dwarfs from the Hipparcos catalog that lie within 0.534° around the ecliptic, roughly corresponding to the unrestricted ETZ. Seven of these G dwarfs are also listed in Table 1,[5] which contains only stars in the rETZ. Observers on planets orbiting the remaining ten G dwarfs in Castellano et al. (2004) would possibly have a harder time identifying Earth as supporting life via transit observations, and hence, their host stars deserve lower priority for SETI.

### *3.2 Galactic stellar model for targets*

Beyond those 82 immediate SETI targets listed in Table 1, how many stars may nowadays[6] host observers of Earth transits? To address this question, we estimate the number of stars that actually reside in the rETZ.

Although M dwarfs are by far the most numerous type of star in the solar neighborhood, numerous issues related to M dwarf habitability have been discussed in the literature, such as tidal locking (Dole, 1964) and the accompanying peril of atmospheric collapse (Joshi e al., 1997), reduced formation of water-rich terrestrial planets (Raymond et al., 2007), weak quiescent ultraviolet radiation from the star (Guo et al., 2010), planetary atmospheric erosion due to weak internal magnetic fields and enhanced stellar winds (Zendejas et al., 2010), forced tidal heating in otherwise habitable moons (Heller, 2012), tidal heating in eccentric terrestrial planets (Barnes et al., 2013), and the possibility of desiccation of terrestrial planets during the pre-main-sequence phase of M dwarfs (Ramirez & Kaltenegger, 2014; Luger & Barnes, 2015). On the other hand, cross-polar circulation may provide a mechanism that prevents atmospheric freeze-out (Haqq-Misra & Kopparapu, 2015), and a stabilizing cloud feedback on terrestrial planets around M dwarfs might even extend the inner edge of the HZ much closer to the star than estimated by 1D longitudinally averaged atmosphere models (Yang et al., 2013). Further, modest and stable tidal heating might actually render terrestrial planets around M dwarfs habitable for trillions of years (Barnes, 2014), and planets around M dwarfs can also stay in the HZ for much longer times than planets around Sun-like stars (Rushby et al., 2013). For a reappraisal of M dwarf habitability see Tarter et al. (2007).

A major issue concerning the existence of observers on planets around FABO stars is related to their relatively short lifetimes compared to the 4.5 billion years required for an intelligent species to arise on Earth. Given the highly stochastic nature of the origin and evolution of life on our home planet, however, it may still be reasonable to assume that a star-gazing life form could evolve on a different planet within, say, a billion years. Hence, F dwarf stars could potentially harbor planets who themselves host observers of Earth transits.

We thus consider a "minimal" model that takes into account KG dwarfs[7], and an "extended" model that invokes MKGF dwarfs. For both models, we assume a vertical stellar density profile

---

[4] http://simbad.u-strasbg.fr/simbad/sim-fsam

[5] These objects have Henry Draper (or HD) designations 108754, 123453, 152956, 165204, 204433, 207921, and 208965.

[6] Given the finite speed of light, "nowadays" here refers to the reader's local time on Earth.

[7] Our minimal model follows up on the idea of superhabitable worlds (Heller and Armstrong, 2014), which suggests that K stars are the most promising host stars for inhabited planets.

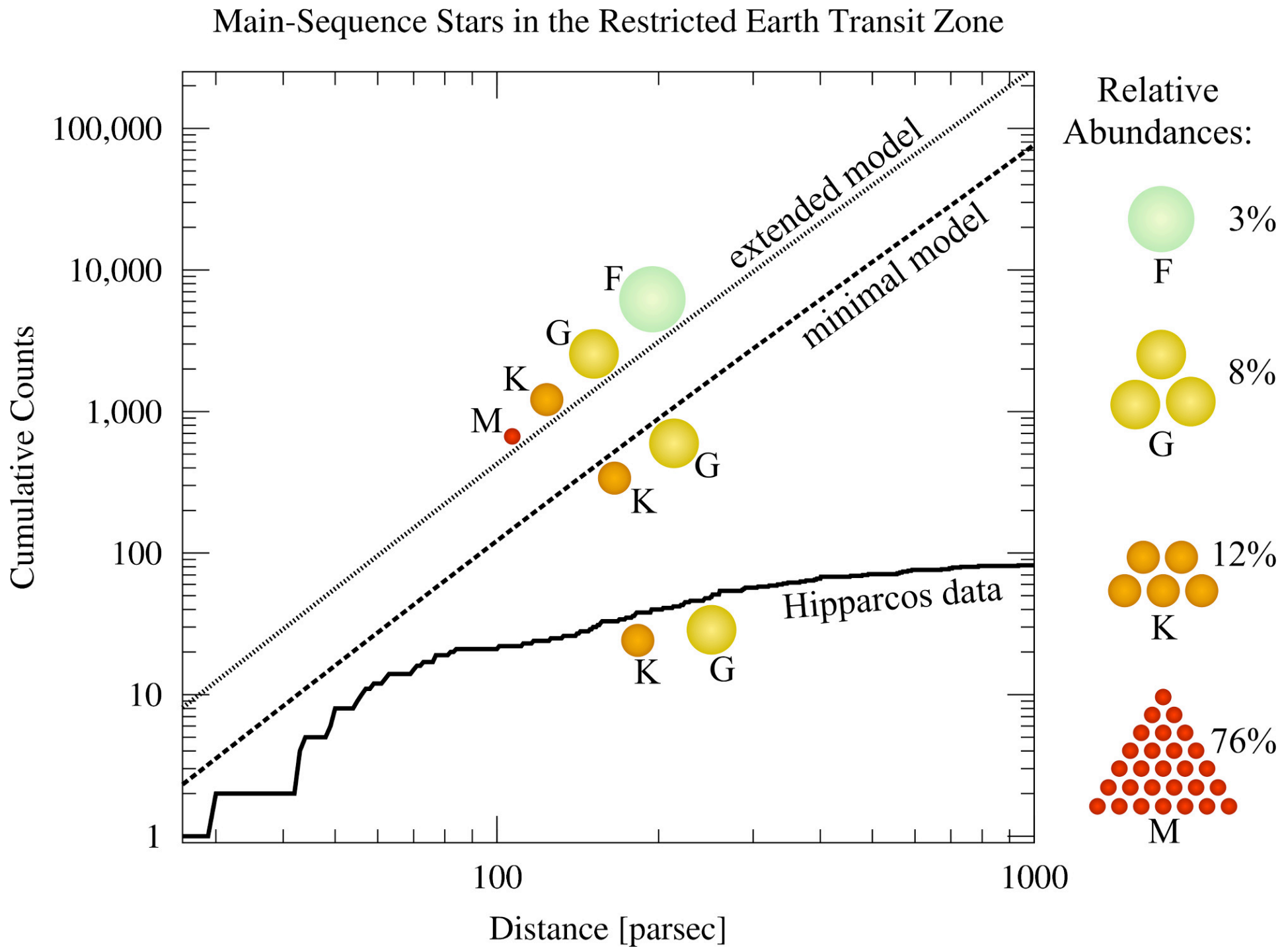

**FIG. 3** Number of main-sequence stars inside the restricted ETZ as a function of distance. The upper dotted line illustrates our disk model (Eq. 6) taking into account MKGF dwarfs, whereas the dashed line considers K and G dwarfs only. According to the latter, almost $10^5$ K and G dwarfs potentially host civilizations, who see Earth transiting the Sun. The bottom solid line is based on Hipparcos K and G dwarf counts (Table 1).

$$\rho(z) = \rho_0 \, \mathrm{e}^{-|z|/z_0} \tag{4}$$

of the Milky Way, where z is the height above the disk midplane, $z_0$ = 300 pc is the local thickness of the thin disk, and $\rho_0 \in \{0.0092, 0.247\}$ $M_\odot$ pc$^{-3}$ is the mass density of KG dwarfs (minimal model) and MKGF dwarfs (extended model), respectively, in the midplane (Flynn et al., 2006). At a distance $r$ from the Sun, the stellar density $\rho(\lambda, r)$ can be described by two contributions, one in the vertical direction and one within the plane. Moving through the ecliptic from geocentric ecliptic longitude $\lambda$ = 0 to 2π (Fig. 2), $\rho(\lambda, r)$ oscillates because the ecliptic is tilted against the galactic plane by $i$ = 63°. Hence,

$$\rho(\lambda', r) = \frac{\rho_0}{2}\Big( [1 - \cos(2\lambda')] + [1 + \cos(2\lambda')] \, \mathrm{e}^{-r\sin(i)/z} \Big) \quad , \tag{5}$$

where $\lambda'$ is chosen to equal zero in the midplane, and we assume that $z_0$ is constant throughout our region of interest.

To obtain the number of KG dwarfs and MKGF inside the rETZ and within a volume of radius $r'$, we integrate $\rho(\lambda', r)$, weigh it with a factor $W = 2.286 \times 10^{-3}$ as per Eq. (2), and divide it by the average stellar mass ($M_{\rm av} \in \{0.65, 0.5\}$ $M_\odot$).

$$\begin{aligned} N_{\rm rETZ}^{\rm KG}(r') &= \frac{W}{M_{\rm av}} \int_0^{r'} \mathrm{d}r \, r^2 \int_0^{2\pi} \mathrm{d}\lambda' \int_0^{\pi} \mathrm{d}\beta \, \sin(\beta) \, \rho(\lambda', r) \\ &= \frac{2\pi\rho_0 W}{M_{\rm av}} \left( \frac{r'^3}{3} + \frac{\mathrm{e}^{-\zeta r'}}{\zeta^3} \Big[ -\zeta r'(\zeta r' + 2) - 2 \Big] + \frac{2}{\zeta^3} \right) \quad , \end{aligned} \tag{6}$$

with $\zeta = \sin(i)/z_0$.

Figure 3 visualizes $N^{\rm KG}{}_{\rm rETZ}(r')$ with a dashed line (minimal model) and $N^{\rm MKGF}{}_{\rm rETZ}(r')$ with a dotted line (extended model). The black solid line is a cumulative KG dwarf star count based on our SIMBAD search (Table 1) and, since Hipparcos measurements are magnitude-limited, almost complete only out to about 80 pc. At larger distances, stars simply become too faint to be detected or well characterized. The relative abundance of stars with MKGF spectral types in the solar neighborhood (Ledrew, 2001) are shown in the legend of Fig. 3. Our disk model predicts 80,000 KG stars within the rETZ and within 1 kpc, whereas a model assuming a constant KG stellar density yields almost twice that amount (not shown). This suggests that the decrease of the stellar density becomes significant towards 1 kpc. With almost $10^5$ KG dwarfs residing as deep as 1 kpc in the ETZ, our minimal model thus predicts about three orders of magnitude more KG dwarfs than are currently known due to observational incompleteness. Our extended model, which takes into account MKGF dwarfs, suggests that almost $3 \times 10^5$ stars could potentially host observers in the rETZ.

Based on recent findings that about 10 % of Sun-like stars host a terrestrial planet in the HZ (Kaltenegger and Sasselov, 2011; Dressing and Charbonneau, 2013; Petigura et al., 2013), our collection of 82 K and G dwarfs in the rETZ implies a total of about eight terrestrial planets around them. But more intriguingly, our total estimate of about $10^5$ KG (or $3 \times 10^5$ MKGF) dwarfs in the rETZ and within 1 kpc suggests that there might be about $10^4$ (or $3 \times 10^4$) terrestrial planets in the HZs around K and G dwarfs and within 1 kpc in the rETZ.

## 4. Discussion

Beyond planets, their yet unseen moons could host life, too (Williams et al., 1997; Heller and Barnes, 2013). Moreover, intelligent civilizations might also travel between stars (Sagan, 1980) and colonize or terraform worlds (planets or moons) that were originally not habitable. Hence, our estimates might be considered lower limits. Ultimately, while there is a certain stellar population inside the ETZ today, stellar proper motions and the variation of the ecliptic cause these targets to float in and out of the ETZ over thousands of years. A star with a proper motion of 100 mas/yr could transition the ETZ within 20 kyr if its motion were perpendicular to the ecliptic. Hence, a substantial population of civilizations may have seen Earth transiting the Sun long ago and identified Earth as supporting life. Although they might not observe our transits today, they might still be trying to contact us. Stars within about a degree to the ETZ thus also serve as potential SETI targets.

Planets orbiting stellar binaries have now been observed (Doyle et al., 2011; Orosz et al., 2012a,b; Welsh et al., 2012; Kostov et al., 2014), and formation models suggest that stellar binarity does not impede planet formation (Martin et al., 2013; Gong et al., 2013). Thus, even if some of the stars listed in Table 1 have low-mass companions, these systems could still host habitable worlds (Forgan, 2014) from a planet formation point of view.

Our study raises immediate science cases for ongoing SETI with the K2 mission (in the optical, Wright et al., 2016), the ATA (radio), the upcoming Square Kilometer Array (SKA, radio), and the Breakthrough Listen Initiatives (radio and optical). With integration times of a few minutes, ATA can detect terrestrial-like radio leakage out to about 80 pc (Blair and ATA Team, 2009). Consequently, it should be capable of tracing intended broadcasts, supposedly emitted with much higher power, out to a few hundred parsecs. On the southern hemisphere, with two operational sites in South Africa and Australia planned to start observations in 2024, SKA will be the largest radio telescope ever. Its extreme sensitivity, combined with a relatively large field of view of about one square degree (Ekers, 2003), will enable scientists to screen a few dozen of our ETZ targets for extraterrestrial beacons. ATA on the northern and SKA on the southern hemisphere have the potential to cover most of the 82 ETZ targets listed in Table 1 in a concerted effort. Ultimately, with their relatively large fields of view, both ATA and SKA can be used to progressively monitor the whole ETZ for intended broadcasts, including $\approx 10^5$ nearby K and G dwarfs not listed in Table 1, on a time scale of days only.

Starting in 2016, the Breakthrough Listen Initiative will combine a radio survey using the 100 m Robert C. Byrd Green Bank Telescope (GBT for short, USA) and the 64 m Parkes Telescope (Australia) with an optical (laser) SETI using the Automated Planet Finder Telescope at Lick Observatory (USA). Four kinds of targets have been announced: (1) $10^6$ nearby stars, (2) the Galactic center, (3) the entire plane of the Milky Way, and (4) 100 nearby galaxies. We propose that Earth's solar transit zone offers a compelling alternative or additional target region for the Breakthrough Listen Initiative. GBT's half-power beamwidth of about 1 arcmin = 1/60 deg at 13 GHz[8] implies a sky-projected area of $\approx$1/3600 deg$^2$ that can be scanned at one time (and one frequency range). With the rETZ covering about $2 \times 10^{-3}$ parts of the entire sky, or roughly 80 deg$^2$ (Eq. 2), this means about $3 \times 10^5$ pointings would be required to fully scan the rETZ with a GBT-like radio telescope. This number is more than three orders of magnitude larger than the number of pointings required to scan the 82 K and G dwarfs listed in Table 1. But it is comparable to the number of pointings implied by a targeted survey of the $10^5$ K and G dwarfs (Fig. 3) that we actually suspect to reside inside the rETZ.

Even if intended messages are absent, the aforementioned radio telescopes could discover leakage

[8] see Proposer's Guide for the Green Bank Telescope as of 17 July 2014 (https://science.nrao.edu/facilities/gbt/proposing/GBTpg.pdf/view)

transmissions. Twenty-two systems in Table 1, and about 100 following our model, are closer than 100 pc. At this proximity, SKA would be capable of detecting even weak leakage signals with a power less than that of some Earth-based Ballistic Missile Early Warning Systems (Loeb and Zaldarriaga, 2007) or with a power comparable to an airport radar as far as 30 pc from Earth (Schilizzi et al., 2007). In particular, at 7.45 (±0.028) pc or 24.31 (±0.09) ly, HR 222 is by far the closest K dwarf in the ETZ.

The ongoing GAIA space mission is supposed to be complete to stars as dim as $V = 20$ mag (Perryman et al., 2001). This limit coincides almost precisely with the apparent visual magnitude of a Sun-like star at 1 kpc. Hence, the early- to mid-G dwarf population in Table 1 and Fig. 3 should grow dramatically from now 37 to about ten thousand objects and then be complete over the whole range of distances considered. Late K dwarfs with visual magnitudes as low as $V = 7$ mag might be complete to only ≈ 100 pc, adding another few hundred targets.

## 5. Conclusions

The region of the Milky Way from where extraterrestrials might observe non-grazing transits of Earth in front of the Sun appears as a band to both sides of the ecliptic with a total width of about 0.528°, assuming a circular Earth orbit. Taking into account Earth's elliptical orbit, we find that the width of the ETZ varies between 0.537° (at geocentric ecliptic longitude $\lambda = 103°$) and 0.520° (at $\lambda = 283°$). Grazing transits can be observed in a band that is about 0.01° wider. Assuming that observers need to see Earth transiting with an impact parameter $b \leq 0.5$ in order to characterize its atmosphere, the width of this restricted ETZ is about 0.262°.

We retrieve 45 K and 37 G dwarf stars within 1 kpc and located inside the rETZ, which implies about eight terrestrial HZ planets based on recent Kepler planet counts. As our sample is severely magnitude limited, we construct a galactic disk model and estimate the true number of K and G dwarfs within 1 kpc and inside the ETZ. This number turns out to be about $10^5$, that is, about $10^3$ times the number of known targets. Hence, our estimate for the number of habitable-zone terrestrial planets in the rETZ goes up to as much as $10^4$. In an extended model that takes into account M and F dwarfs in the rETZ, counts increase by another factor of three.

We have literally adopted Sagan's vision of Earth as "a mote of dust suspended in a sunbeam" (Sagan, 1994): our planet may be seen by distant observers against our Sun's light, allowing them to detect us. As an ultimate consequence, even if our species chose to remain radio-quiet to eschew interstellar contact, we cannot hide from observers located in Earth's solar transit zone, if they exist. Table 1 lists 82 host stars of these potential observers and may serve as an initial priority target list for SETI observations in Earth's transit zone. A wider survey of the hundreds of thousands of targets predicted by our galactic model constitutes an effective strategy for the Breakthrough Listen Initiative and other SETI observations to search for extraterrestrial calling partners.

## Acknowledgements

We thank two anonymous referees and Jason Wright for their constructive reports on the initial version of this manuscript. R. Heller has been supported by the Origins Institute at McMaster University, by the Canadian Astrobiology Program (a Collaborative Research and Training Experience Program funded by the Natural Sciences and Engineering Research Council of Canada, NSERC), by the Institute for Astrophysics Göttingen, and by a Fellowship of the German Academic Exchange Service (DAAD). R. E. Pudritz is supported by a Discovery grant from NSERC. He also thanks the Max Planck Institute for Astronomy, Heidelberg, and the Center for Astronomy – Institute for Theoretical Astrophysics at Heidelberg University for their hospitality and support of his sabbatical leave during which the manuscript was finalized. This work has made use of NASA's Astrophysics Data System Bibliographic Services and of the SIMBAD database, operated at CDS, Strasbourg, France. Discussions with Dimitris Mislis in the frame of the DFG-funded Graduiertenkolleg 1351 "Extrasolar Planets and Their Host Stars" initiated this study many years ago.

Table 1: K dwarf stars (45 by numbers) and G dwarf stars (37) within 1 kpc from the Sun and inside the restricted ETZ. Objects are sorted by increasing distances.

| Identifier | ICRS coordinates J2000/2000 | Ecliptic coordinates J2000/2000 | Proper Motion mas/yr | Parallax mas | Distance pc | $V$ mag. | Spectral Type |
|---|---|---|---|---|---|---|---|
| HR 222 | 00 48 22.97665 +05 16 50.2129 | 013.1821587 +00.0825339 | 757.11 -1141.33 | 134.14 (±0.51) | $7.45^{+0.03}_{-0.03}$ | 5.74 | K2.5V |
| GJ 757 | 19 24 34.23191 -22 03 43.8557 | 289.5287641 -00.0448508 | -236.43 -454.54 | 34.11 (±2.11) | $29.32^{+1.93}_{-1.71}$ | 10.9 | K8Vk: |
| HD 53532 | 07 06 16.80335 +22 41 00.5537 | 105.2557439 +00.1174720 | -89.98 -78.56 | 23.60 (±0.82) | $42.37^{+1.53}_{-1.42}$ | 8.27 | G0V |
| HD 115153 | 13 15 30.76740 -08 03 18.5701 | 200.4670388 -00.0647171 | 42.18 58.71 | 23.76 (±0.67) | $42.09^{+1.22}_{-1.15}$ | 8.06 | G5 |
| LTT 11954 | 06 57 46.33755 +22 53 33.1761 | 103.2837246 +00.1168203 | -142.85 -142.91 | 23.12 (±0.66) | $43.25^{+1.27}_{-1.20}$ | 7.63 | G0 |
| HD 108754 | 12 29 42.73517 -03 19 58.6869 | 188.1369160 -00.1147546 | -327.65 -562.26 | 20.56 (±0.99) | $48.64^{+2.46}_{-2.23}$ | 9.03 | G7V |
| G 3-38 | 02 03 33.03157 +12 35 04.9890 | 033.1168429 +00.0348568 | 381.09 -51.86 | 20.30 (±3.71) | $49.26^{+11.02}_{-7.61}$ | 11.54 | K7 |
| HD 20477 | 03 18 19.98264 +18 10 17.8452 | 051.9747902 -00.0924300 | -84.56 -103.48 | 20.16 (±0.64) | $49.60^{+1.63}_{-1.53}$ | 7.54 | G0 |
| HD 78451 | 09 08 49.27841 +16 25 03.5462 | 134.7386517 +00.0053713 | -40.57 -71.38 | 18.21 (±2.91) | $54.91^{+10.44}_{-7.57}$ | 9.47 | G5 |
| StKM 2-1619 | 22 31 18.99284 -09 25 34.7527 | 336.0020753 -00.1250324 | -141.48 -195.84 | 17.90 (±2.60) | $55.87^{+9.49}_{-7.09}$ | 11.15 | K5V |
| LTT 6827 | 17 05 59.62335 -22 51 24.3597 | 257.5762891 +00.0021104 | 34.76 -325.61 | 17.55 (±2.03) | $56.98^{+7.45}_{-5.91}$ | 10.0 | G |
| NLTT 20353 | 08 50 36.87345 +17 41 21.5587 | 130.2052944 +00.0036610 | -149.74 -59.67 | 17.07 (±2.27) | $58.58^{+8.99}_{-6.88}$ | 9.323 | K0 |
| G 156-19 | 22 33 21.52793 -09 03 48.8063 | 336.6043076 +00.0261599 | 298.54 -62.00 | 16.17 (±0.96) | $61.84^{+3.90}_{-3.47}$ | 8.71 | G0 |
| HD 123453 | 14 08 04.09936 -12 55 42.1319 | 214.2696634 +00.0156760 | 116.89 -110.14 | 15.90 (±1.39) | $62.89^{+6.03}_{-5.06}$ | 7.95 | G9V:+... |
| BD+18 487 | 03 27 55.30483 +18 52 56.4159 | 054.3533816 +00.0238173 | 13.83 -59.44 | 14.41 (±1.43) | $69.40^{+7.65}_{-6.27}$ | 8.48 | G0 |
| HD 224448 | 23 58 04.536 -00 07 41.49 | 359.507608 +00.073754 | -45.28 -19.25 | 14.13 (±1.25) | $70.77^{+6.87}_{-5.76}$ | 9.01 | G0 |
| LTT 8911 | 22 12 50.82036 -10 55 31.3525 | 331.2230435 +00.1223774 | 178.99 -183.40 | 13.85 (±2.09) | $72.20^{+12.83}_{-9.47}$ | 10.8 | K3 |
| LTT 7183 | 18 05 46.73402 -23 31 03.8191 | 271.3247005 -00.0850796 | 25.24 -212.03 | 13.13 (±1.18) | $76.16^{+7.52}_{-6.28}$ | 9.05 | G7V |
| LTT 7494 | 18 54 12.798 -22 54 25.32 | 282.466510 -00.052375 | -161.7 -368.0 | 13.00 (±7.00) | $76.92^{+89.74}_{-26.92}$ | 8.48 | G5 |
| HD 81563 | 09 26 49.40417 +14 55 40.7481 | 139.3192061 -00.1051346 | 21.47 -91.74 | 12.22 (±0.92) | $81.83^{+6.66}_{-5.73}$ | 8.27 | G0 |
| HD 121865 | 13 58 26.84077 -12 03 33.4252 | 211.7657893 +00.0303109 | 133.80 -150.88 | 11.91 (±0.88) | $83.96^{+6.70}_{-5.78}$ | 7.05 | G5 |
| HD 204433 | 21 28 50.03197 -14 49 34.9337 | 319.8123553 +00.0494591 | -21.87 -100.07 | 9.94 (±1.37) | $100.60^{+16.08}_{-12.19}$ | 9.1 | G2V |
| HD 284145 | 04 04 56.52494 +20 51 23.3074 | 063.2768930 +00.0460690 | 10.54 -53.70 | 9.00 (±1.38) | $111.11^{+20.12}_{-14.77}$ | 8.98 | G0 |
| HD 109376 | 12 34 13.27236 -03 43 16.1748 | 189.3239649 -00.0283740 | 14.16 -57.19 | 8.58 (±0.86) | $116.55^{+12.98}_{-10.62}$ | 8.57 | G5 |
| HD 244354 | 05 31 04.38928 +23 12 34.7855 | 083.3563104 -00.0630026 | 10.45 -38.04 | 7.95 (±1.29) | $125.79^{+24.34}_{-17.56}$ | 9.11 | G0 |
| HD 152956 | 16 57 30.23696 -22 38 37.1391 | 255.6074904 +00.0183989 | -0.12 -29.91 | 7.53 (±1.42) | $132.80^{+30.86}_{-21.07}$ | 9.78 | G0V |
| HD 117004 | 13 27 29.65820 -09 11 33.7450 | 203.6390818 -00.0160648 | -55.67 -12.36 | 7.13 (±1.42) | $140.25^{+34.88}_{-23.29}$ | 9.48 | G0 |
| HD 207921 | 21 52 54.70756 -12 58 41.6150 | 325.9382627 -00.1118968 | -35.35 -79.37 | 7.02 (±0.92) | $142.45^{+21.48}_{-16.51}$ | 8.77 | G3V |
| HD 15002 | 02 25 27.73995 +14 32 35.7739 | 038.7897351 +00.1198081 | 20.82 -49.85 | 6.74 (±1.13) | $148.37^{+29.89}_{-21.30}$ | 9.19 | G0 |
| HD 31363 | 04 56 06.63500 +22 34 35.6076 | 075.2805471 -00.0502722 | 38.65 -53.50 | 6.64 (±0.86) | $150.60^{+22.41}_{-17.27}$ | 7.17 | K0 |
| HD 218553 | 23 09 01.53846 -05 20 30.5894 | 346.1996508 +00.1116472 | 16.63 2.73 | 6.51 (±1.10) | $153.61^{+31.23}_{-22.20}$ | 8.61 | K0 |
| HD 210241 | 22 09 18.13872 -11 21 58.3422 | 330.2538687 +00.0175564 | 4.42 -42.40 | 6.43 (±0.73) | $155.52^{+19.92}_{-15.86}$ | 7.86 | K0 |
| HD 110918 | 12 45 32.02766 -04 48 39.3989 | 192.3457381 +00.0737266 | -18.62 7.16 | 6.39 (±0.56) | $156.50^{+15.03}_{-12.61}$ | 7.16 | K2 |
| HD 106352 | 12 14 10.58599 -01 38 31.7169 | 183.9043926 -00.0981432 | 92.45 -98.42 | 5.94 (±1.42) | $168.35^{+52.89}_{-32.48}$ | 9.71 | G5 |
| LTT 8795 | 22 00 13.00098 -12 13 29.6524 | 327.8705876 -00.0123927 | 169.20 -93.07 | 5.75 (±1.35) | $173.91^{+53.36}_{-33.07}$ | 9.23 | G5V |
| BD-22 4248 | 16 56 41.05634 -22 40 27.2304 | 255.4227874 -00.0323920 | -11.85 -2.97 | 5.61 (±2.70) | $178.25^{+165.39}_{-57.92}$ | 9.84 | G5 |
| HD 14243 | 02 18 34.05265 +13 57 00.9562 | 037.0170514 +00.0997387 | 6.36 -14.36 | 5.54 (±1.14) | $180.51^{+46.77}_{-30.81}$ | 8.44 | K0 |
| HD 71907 | 08 30 31.01625 +18 56 34.7115 | 125.2737250 -00.0074509 | -13.48 -11.24 | 5.50 (±1.13) | $181.82^{+47.02}_{-30.99}$ | 8.63 | K2 |
| HD 3819 | 00 40 49.92047 +04 28 41.9361 | 011.1373198 +00.0778609 | -11.92 -20.75 | 5.16 (±0.72) | $193.80^{+31.43}_{-23.73}$ | 7.97 | G5 |
| HD 66287 | 08 03 34.13418 +20 20 18.6938 | 118.7779183 -00.0675791 | 13.93 -27.38 | 5.16 (±0.66) | $193.80^{+28.42}_{-21.98}$ | 7.03 | G5 |
| HD 61660 | 07 41 12.69882 +21 26 48.8100 | 113.4412891 +00.0431439 | 11.69 -42.56 | 4.91 (±0.91) | $203.67^{+46.33}_{-31.85}$ | 7.87 | K2 |
| G 14-32 | 13 08 25.78882 -07 18 30.4548 | 198.5601373 -00.0373871 | -219.83 82.11 | 4.71 (±1.03) | $212.31^{+59.43}_{-38.01}$ | 8.35 | G5 |
| HD 218399 | 23 07 50.70029 -05 41 51.5236 | 345.7908778 -00.1024129 | 20.51 -3.66 | 4.55 (±1.34) | $219.78^{+91.75}_{-50.00}$ | 8.54 | K0 |
| 44 Cnc | 08 43 08.35527 +18 09 01.9705 | 128.3684830 -00.0222521 | -1.25 -3.04 | 4.49 (±0.82) | $222.72^{+49.76}_{-34.39}$ | 8.03 | K0 |
| 72 Aqr | 22 50 46.30141 -07 18 42.9276 | 341.2553953 +00.0346433 | -25.71 -14.01 | 4.48 (±0.65) | $223.21^{+37.89}_{-28.29}$ | 7.0 | K0 |
| HD 16178 | 02 36 11.96153 +15 08 55.3242 | 041.4430635 -00.1219804 | -1.98 -40.02 | 4.38 (±0.74) | $228.31^{+46.42}_{-33.00}$ | 8.4 | K0 |
| HD 72115 | 08 31 41.29613 +18 59 15.8392 | 125.5314085 +00.1034379 | -26.31 -20.99 | 4.12 (±0.51) | $242.72^{+34.29}_{-26.74}$ | 6.45 | K0 |
| HD 218846 | 23 11 17.17017 -05 05 55.5145 | 346.8126666 +00.1170492 | 15.77 4.11 | 4.10 (±1.04) | $243.90^{+82.90}_{-49.36}$ | 8.85 | K0 |
| HD 2690 | 00 30 31.24132 +03 16 00.7985 | 008.2962799 -00.0255395 | 16.51 -0.46 | 4.01 (±0.98) | $249.38^{+80.66}_{-48.98}$ | 8.44 | K2 |
| HD 940 | 00 13 47.57526 +01 23 00.9161 | 003.7137699 -00.1011061 | -25.67 -14.22 | 3.97 (±0.61) | $251.89^{+45.73}_{-33.55}$ | 6.77 | K0 |
| HD 17795 | 02 51 52.36942 +16 29 51.8022 | 045.4438836 +00.0334572 | -3.84 -15.93 | 3.97 (±0.93) | $251.89^{+77.06}_{-47.81}$ | 8.13 | K5 |
| HD 5214 | 00 53 58.22129 +05 48 32.9485 | 014.6686315 +00.0303990 | 27.17 3.95 | 3.86 (±0.57) | $259.07^{+44.88}_{-33.33}$ | 7.92 | G5 |
| HD 50318 | 06 53 36.58256 +22 46 34.2676 | 102.3401752 -00.0908021 | -0.68 -8.08 | 3.86 (±0.86) | $259.07^{+74.27}_{-47.20}$ | 8.22 | K0 |
| HD 29097 | 04 35 36.74097 +22 01 14.4366 | 070.5072283 -00.0023000 | 16.31 -24.59 | 3.85 (±1.46) | $259.07^{+158.67}_{-71.42}$ | 8.21 | K0 |
| BD+20 1979 | 08 01 42.33483 +20 26 49.9871 | 118.3283130 -00.0497126 | -7.39 -10.44 | 3.50 (±1.08) | $285.71^{+127.51}_{-67.37}$ | 9.05 | G5 |
| HD 54427 | 07 09 44.41414 +22 21 52.5414 | 106.0864687 -00.1063125 | 17.69 -12.19 | 3.48 (±0.68) | $287.36^{+69.79}_{-46.97}$ | 7.56 | K2 |
| HD 118866 | 13 39 43.95528 -10 14 51.5276 | 206.8316729 +00.1025172 | -22.75 9.18 | 3.46 (±0.93) | $289.02^{+106.24}_{-61.23}$ | 8.44 | K2 |
| HD 39184 | 05 51 50.65689 +23 22 58.0589 | 088.1285815 -00.0432527 | -8.36 -20.44 | 3.27 (±0.72) | $305.81^{+86.35}_{-55.19}$ | 6.98 | K5 |
| BD+18 472 | 03 23 39.55279 +18 33 20.7050 | 053.2946619 -00.0423735 | -3.38 -11.71 | 3.12 (±1.66) | $320.51^{+364.42}_{-111.31}$ | 8.93 | K0 |
| HD 7990 | 01 19 28.31292 +08 23 46.9468 | 021.5030840 +00.0134072 | -14.37 34.86 | 3.01 (±0.65) | $332.23^{+91.50}_{-59.00}$ | 7.91 | K0 |
| HD 118777 | 13 39 07.04823 -10 23 24.7498 | 206.7420336 -00.0850406 | 9.77 -22.94 | 2.94 (±1.43) | $340.14^{+322.12}_{-111.30}$ | 8.47 | K0 |
| BD-13 6044 | 21 53 49.08240 -12 54 16.5428 | 326.1709588 -00.1173764 | -11.92 -35.21 | 2.88 (±1.83) | $347.22^{+605.16}_{-134.91}$ | 10.13 | G0 |
| BD+15 2068 | 09 32 57.78706 +14 24 26.9151 | 140.8924859 -00.1306424 | 35.74 -46.11 | 2.73 (±1.27) | $366.30^{+318.63}_{-116.30}$ | 9.25 | K0 |
| HD 70842 | 08 24 49.68026 +19 10 13.0292 | 123.9140629 -00.1079141 | 1.97 -3.67 | 2.66 (±1.19) | $375.94^{+304.33}_{-116.20}$ | 8.91 | K0 |
| HD 90762 | 10 28 55.89407 +09 33 54.3328 | 155.4041598 +00.0379253 | 14.48 -30.82 | 2.57 (±0.85) | $389.11^{+192.29}_{-96.71}$ | 8.44 | G5 |
| HD 88680 | 10 13 50.38192 +10 50 31.3131 | 151.4804623 -00.1136457 | 5.36 4.87 | 2.54 (±0.71) | $393.70^{+152.75}_{-86.01}$ | 8.02 | G5 |
| BD+04 122 | 00 48 02.45463 +05 06 12.7318 | 013.0348238 -00.0474695 | 0.11 -0.90 | 2.50 (±1.02) | $400.00^{+275.68}_{-115.91}$ | 8.99 | K2 |
| HD 121185 | 13 54 07.92284 -11 41 43.3350 | 210.6494013 +00.0044963 | 0.12 18.81 | 2.50 (±0.69) | $400.00^{+152.49}_{-86.52}$ | 7.37 | K2 |
| HD 63105 | 07 47 55.89599 +21 02 05.5168 | 115.0551127 -00.0880687 | -12.51 -8.90 | 2.20 (±0.67) | $454.55^{+199.05}_{-106.11}$ | 7.78 | K2 |
| HD 81541 | 09 26 40.48560 +15 06 57.9390 | 139.2263973 +00.0623751 | -13.49 -2.81 | 2.09 (±1.26) | $478.47^{+726.35}_{-170.96}$ | 9.19 | G |
| HD 3082 | 00 34 07.51580 +03 48 32.0706 | 009.3364777 +00.1184635 | 4.52 2.76 | 2.04 (±1.23) | $490.20^{+744.37}_{-184.39}$ | 8.86 | K2 |
| HD 46279 | 06 33 27.26922 +23 10 43.3360 | 097.6842660 -00.0377662 | 0.08 -4.62 | 1.80 (±1.18) | $555.56^{+1057.35}_{-219.99}$ | 8.37 | K0 |
| BD+15 391 | 02 49 04.16610 +16 07 22.0335 | 044.6909552 -00.1281619 | -16.32 -1.78 | 1.78 (±1.23) | $561.80^{+1256.38}_{-229.57}$ | 9.21 | K0 |
| HD 222294 | 23 39 23.37144 -02 09 16.4705 | 354.4165961 +00.0690674 | 5.02 -1.40 | 1.76 (±1.22) | $568.18^{+1283.67}_{-232.61}$ | 8.64 | K5 |
| HD 116210 | 13 22 18.76319 -08 41 58.5069 | 202.2678209 -00.0325619 | -2.02 6.12 | 1.72 (±0.94) | $581.40^{+700.66}_{-205.46}$ | 8.57 | G5 |
| HD 10917 | 01 47 21.58819 +10 57 13.2703 | 028.8344633 -00.1139321 | 6.56 -2.02 | 1.69 (±0.97) | $591.72^{+797.17}_{-215.78}$ | 8.77 | K2 |
| HD 102105 | 11 45 05.53604 +01 34 43.3932 | 175.9530492 -00.0326170 | -7.50 9.64 | 1.49 (±1.01) | $671.14^{+1412.19}_{-271.14}$ | 8.59 | K5 |
| HD 107136 | 12 19 07.46767 -02 02 24.8982 | 185.1972389 +00.0269111 | -27.53 5.57 | 1.44 (±1.27) | $694.44^{+5187.91}_{-325.44}$ | 9.24 | K2 |
| HD 106351 | 12 14 09.77273 -01 35 26.8989 | 183.8809015 -00.0523691 | -38.18 10.20 | 1.42 (±1.07) | $704.23^{+2152.92}_{-302.62}$ | 9.05 | G5 |
| HD 90225 | 10 25 02.00411 +09 58 56.3018 | 154.3577234 +00.0752950 | -16.81 -20.19 | 1.37 (±0.84) | $729.93^{+1156.87}_{-277.44}$ | 8.25 | K0 |
| HD 64768 | 07 56 15.07077 +20 49 03.0689 | 117.0041988 +00.0614611 | 3.88 -14.67 | 1.27 (±0.95) | $787.40^{+2337.60}_{-336.95}$ | 8.32 | K0 |
| HD 216905 | 22 57 04.57976 -06 48 56.5140 | 342.8906179 -00.1031304 | 17.41 -7.73 | 1.07 (±1.46) | $934.58^{+nan}_{-539.32}$ | 9.35 | K0 |